\documentclass[aps,prd,reprint,superscriptaddress,nofootinbib, longbibliography]{revtex4-1}

\usepackage{amsmath,amssymb, amsthm,amstext}
\usepackage{natbib}
\usepackage{graphicx}
\usepackage{color}
\usepackage{array, enumerate}
\usepackage{bm}
\usepackage{multirow}
\usepackage[breaklinks,colorlinks,citecolor=blue]{hyperref}
\usepackage{braket}
\usepackage{txfonts}
\usepackage[normalem]{ulem}
\usepackage{orcidlink}

\def\be{\begin{equation}}
\def\ee{\end{equation}}

\def\apjl{Astroph.J.Lett.}

\def\mnras{MNRAS}

\def\nar{New Astronomy Reviews}
\def\aap{Astronomy \& Astrophysics}

\begin{document}
\title{Gravitational waves associated with the {\em r}-mode instability from neutron star-white dwarf mergers}

\author{Shu-Qing Zhong~\orcidlink{0000-0002-1766-6947}}
\email{sq\_zhong@qq.com}
\affiliation{School of Science, Guangxi University of Science and Technology, Liuzhou 545006, China}

\author{Yan-Zhi Meng~\orcidlink{0000-0002-1122-1146}}
\affiliation{School of Science, Guangxi University of Science and Technology, Liuzhou 545006, China}

\author{Jia-Hong Gu}
\affiliation{School of Science, Guangxi University of Science and Technology, Liuzhou 545006, China}

\begin{abstract}
Neutron star-white dwarf (NS-WD) binaries evolve into either ultra-compact X-ray binaries undergoing stable mass transfer or direct mergers by unstable mass transfer.
While much attention has been on gravitational wave (GW) emissions from NS-WD binaries with the former evolutionary pathway, 
this work explores GW emissions related to {\em r}-mode instability of the accreting NSs in NS-WD mergers particularly with WD's mass $\gtrsim 1M_{\odot}$. 
Due to considerably high accretion rates, the GW emissions associated with both {\em r}-modes and magnetic deformation intrinsically induced by {\em r}-modes presented in this work are much stronger than those in NS-WD binaries categorized as intermediate-mass or low-mass X-ray binaries, rendering them interesting sources for the advanced Laser Interferometer Gravitational Wave Observatory and upcoming Einstein Telescope.
Moreover, these strong GW emissions might accompany some intriguing electromagnetic emissions such as peculiar long gamma-ray bursts (LGRBs), fast blue optical transients including kilonova-like emissions associated with peculiar LGRBs, and/or fast radio bursts. 
\end{abstract}
\maketitle

\section{Introduction}
\label{sec:intro}
The fate of a neutron star-white dwarf (NS-WD) binary hinges on its WD mass $M_{\rm WD}$ and the critical WD mass $M_{\rm WD,crit}$, as established by seminal works \cite{hje87,hur02}. Depending on the comparison between $M_{\rm WD}$ and $M_{\rm WD,crit}$, the binary has two evolutionary pathways. The first one is that it evolves into an ultra-compact X-ray binary through long-term stable mass transfer ($M_{\rm WD}<M_{\rm WD,crit}$). The second is that it undergoes a direct rapid merger due to unstable mass transfer ($M_{\rm WD}>M_{\rm WD,crit}$).

Presently, the predominant focus in gravitational wave (GW) studies of NS-WD binaries centers on the former evolutionary pathway. On one hand, the GW emissions arising from the orbital motion of such systems typically fall within the low frequency range of $\sim10^{-4}-1$ Hz prior to contact \cite{yin23,kang24} or post contact \cite{tau18,yu21}, rendering them intriguing targets for upcoming missions like the Laser Interferometer Space Antenna (LISA; \cite{nel09,nel10}), Taiji \cite{ruan20}, and Tianqin \cite{wang19}. On the other hand, post-contact GW emissions associated with the excitation of Rossby oscillations ({\em r}-modes) and/or magnetic deformation induced by {\em r}-modes of the rotating NS due to the accretion from the WD typically occur in the high frequency range of $\sim100-1000$ Hz \cite{yos00,cuo12}, 
making them compelling targets for the advanced Laser Interferometer Gravitational Wave Observatory 
(aLIGO; \cite{har10}) and Einstein Telescope (ET; \cite{hild11}).
For example, several authors have delved into GWs stemming from {\em r}-mode instability in NS-WD systems that are typically categorized as intermediate-mass or low-mass X-ray binaries (IMXBs or LMXBs) undergoing stable mass transfer \cite{levin99,and99,cuo12,mah13}. 

On the contrary, for NS-WD systems coming into the second evolutionary pathway, their GWs specially for those GWs associated with {\em r}-mode instability have been little studied. Therefore, our work aims to investigate the GWs associated with {\em r}-mode instability from the accreting NSs in those NS-WD systems that undergo unstable mass transfer and thus enter a rapid merger phase. Due to the heightened accretion rates in such systems, the GW signatures relative to {\em r}-mode instability and magnetic deformation induced by {\em r}-mode instability are expected to be more pronounced, compared to the corresponding GW signals in the systems with the first evolutionary pathway.

The paper is structured as follows: In Section \ref{sec:evolution}, we elucidate the evolution of {\em r}-mode instability in the accreting NS within an NS-WD merger, taking into account the instability growth driven by the GW radiation via the Chandrasekhar-Friedman-Schutz (CFS) mechanism \cite{chan70,frie78} and the suppression due to both the viscosity and the internal toroidal magnetic field intrinsically induced by {\em r}-modes. Section \ref{sec:numerical} presents our numerical calculations pertaining to the evolution of {\em r}-mode instability and showcases the results. Section \ref{sec:gw} delves into the GW emissions relevant to {\em r}-modes and magnetic deformation intrinsically induced by {\em r}-modes. 
Section \ref{sec:counterparts} gives a discussion on possible electromagnetic (EM) counterparts.  
Finally, Section \ref{sec:summary} provides a summary.

\section{Evolution of {\em R}-mode Instability}\label{sec:evolution}
In general, the {\em r}-modes of rotating barotropic Newtonian stars are solutions of the perturbed fluid equations having Eulerian velocity perturbations. The evolution of these modes is influenced by both dissipative dampings and driving effects. Dissipative dampings usually include viscosity \cite[e.g.,][]{lind98} and magnetic dampings \cite{cuo10}. While driving effects mainly contain the GW radiation via the CFS mechanism. In this work, we consider the {\em r}-modes of the accreting NS in an NS-WD merger system. Similar to the {\em r}-mode instability of the accreting NS in a LMXB \cite{cuo12} or in a core-collapse supernova \cite{wang17} 
which is driven by the GW radiation via the CFS mechanism and suppressed by the viscosity and toroidal magnetic field, 
the evolution of the {\em r}-mode instability in the accreting NS within an NS$-$WD merger will be governed by following four differential equations \cite{cuo12,wang17}
\begin{equation}\label{eq:alpha}
	\begin{aligned}
		\frac{1}{\alpha} \frac{d\alpha}{dt}= & \frac{\dot{M} A_{-}}{2 M A_{+}}-\frac{1}{A_{+}}\left(\frac{1}{t_{\mathrm{acc}}}+\frac{1}{t_{\mathrm{dip}}}+\frac{1}{t_{B, \mathrm{gw}}}\right) \\
		& -\frac{A_{-}}{A_{+}}\left(\frac{1}{t_{\mathrm{sv}}}+\frac{1}{t_{\mathrm{bv}}}+\frac{1}{t_{B, \mathrm{t}}}\right)-\frac{1}{t_{\mathrm{gw}}}, 
	\end{aligned}
\end{equation}
\begin{equation}\label{eq:Omega}
	\begin{aligned}
		\frac{1}{\Omega} \frac{d\Omega}{dt}= & -\frac{\dot{M}}{A_{+} M}+\frac{1}{A_{+}}\left(\frac{1}{t_{\mathrm{acc}}}+\frac{1}{t_{\mathrm{dip}}}+\frac{1}{t_{B, \mathrm{gw}}}\right) \\
		& -\frac{3 \alpha^2 \tilde{J}}{\tilde{I} A_{+}}\left(\frac{1}{t_{\mathrm{sv}}}+\frac{1}{t_{\mathrm{bv}}}+\frac{1}{t_{B, \mathrm{t}}}\right),
	\end{aligned}
\end{equation}

\begin{equation} \label{eq:B_t}
	\frac{dB_{\mathrm{t}}}{dt} \approx\left(\frac{4}{3 \pi}\right)^{1 / 2} B \alpha^2 \Omega,
\end{equation}
and 
\begin{equation}\label{eq:T}
	\frac{d}{dt}\left[\frac{1}{2}C_{\rm V} T\right]=\dot{\epsilon_{\rm s}} + \dot{\epsilon_{\rm n}} - \dot{\epsilon_{\nu}}.
\end{equation}
Here $\alpha$, $\Omega$, $B_{\rm t}$, and $T$ represent the {\em r}-mode amplitude, spin angular frequency, volume-averaged toroidal magnetic field, 
and spatially averaged temperature of the NS, respectively. We will introduce each term within above differential equations in detail below.

First, the NS mass $M$ in Equations (\ref{eq:alpha}) and (\ref{eq:Omega}) varies over time $t$ and relates to the accretion rate $\dot{M}$ by
\begin{equation}\label{eq:M}
	M(t)=\begin{cases} M_{\rm i}+\int_0^t \dot{M} dt, t\leqslant t_{\rm c}~({\rm accretion~phase})      \\
		M_{\rm i}+\int_0^{t_{\rm c}}\dot{M} dt, t>t_{\rm c}~({\rm propeller~phase}),
		\end{cases}
\end{equation}
where $M_{\rm i}$ denotes the initial mass and $t_{\rm c}$ marks the transition time from the accretion phase to the propeller phase. 
This transition occurs once the relation between the magnetospheric radius ($r_{\rm m}$) and the co-rotation radius ($r_{\rm c}$) 
shifts from $r_{\rm m}\leqslant r_{\rm c}$ to $r_{\rm m}> r_{\rm c}$ due to the evolving accretion rate. 
Here the magnetospheric radius is written by
\be \label{eq:r_m}
r_{\rm m}=\left(\frac{\mu^4}{G M \dot{M}^2}\right)^{1/7}
\ee
and the co-rotation radius is
\be \label{eq:r_c}
r_{\rm c}=\left(\frac{G M}{\Omega^2}\right)^{1/3},
\ee
where $\mu=BR^3$ represents the magnetic dipole moment with $B$ as the surface magnetic field strength.
The accretion rate $\dot{M}$ is modeled by a polynomial function for fitting numerical results\footnote{In the figure 3 of Ref. \cite{kal23}, the depicted accretion rate corresponds to an NS-WD system with a mass ratio of 1.25$M_{\odot}$:1.0$M_{\odot}$. However, in our study, we consider an NS-WD system with a mass ratio of 1.4$M_{\odot}$:1.0$M_{\odot}$. Despite this difference in mass ratios, the accretion rate we utilize is comparable to that of the 1.25$M_{\odot}$:1.0$M_{\odot}$ system, as evidenced by comparisons of peak accretion rates in the tables 2, 3, and 4 of Ref. \cite{kal23}.} in the figure 3 of Ref. \cite{kal23}
\begin{widetext}
\begin{eqnarray}
	\dot{M}\simeq\begin{cases}10^{-1.34+0.00167\times {\rm log}_{10}(t)-1.025\times [{\rm log}_{10}(t)]^2+0.64\times [{\rm log}_{10}(t)]^3-0.47\times [{\rm log}_{10}(t)]^4}\ M_{\odot}{\rm s^{-1}}, {\rm CEEW} \\
	10^{-2.75+0.39\times {\rm log}_{10}(t)-0.74\times [{\rm log}_{10}(t)]^2+0.018\times [{\rm log}_{10}(t)]^3}\ M_{\odot}{\rm s^{-1}}, {\rm HENW},
	\end{cases}
	\label{eq:M_dot}
\end{eqnarray}
\end{widetext}
where $t$ is measured in seconds, CEEW and HENW separately represent the constant entropy efficient wind 
and high-entropy normal wind prescriptions for the WD debris disk \cite{kal23}.

\begin{figure*}
	\includegraphics[scale=0.43]{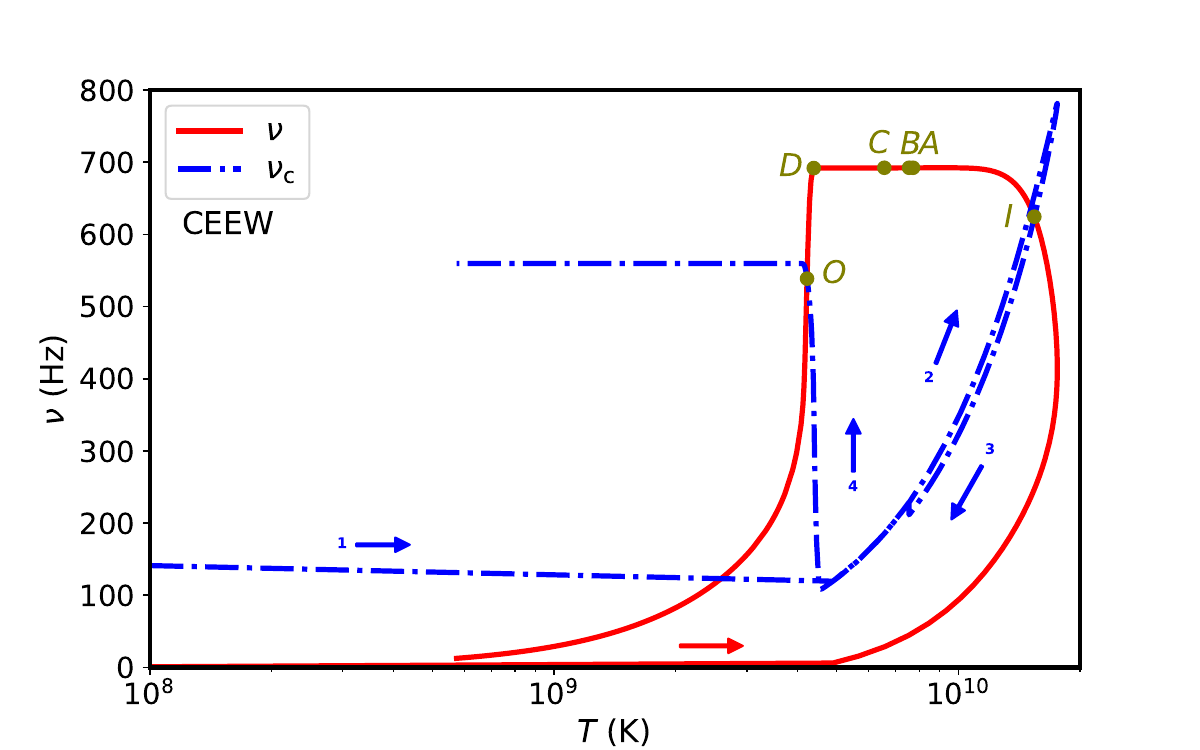}
	\includegraphics[scale=0.43]{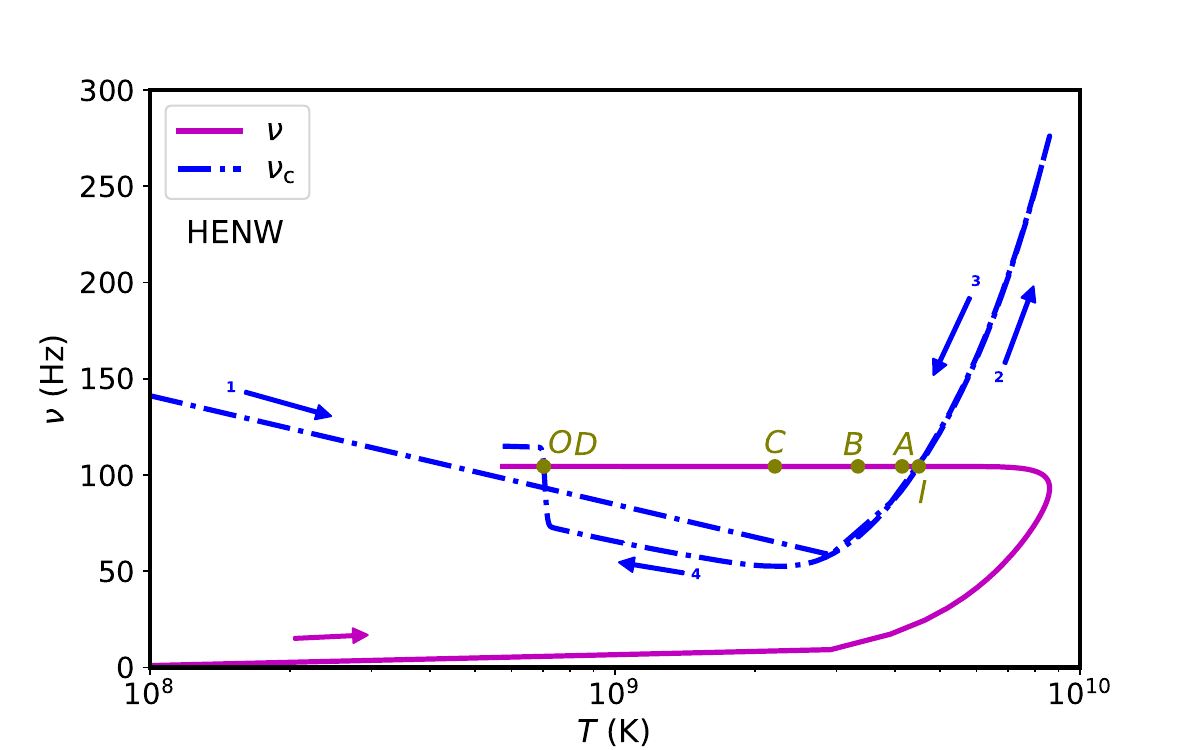}
	\caption{The solid lines represent evolution as a function of spin frequency $\nu$ and temperature $T$ for the accreting NS in an NS-WD merger. {\em R}-mode instability window is determined by the critical frequency $\nu_{\rm c}$ denoted by the dashed dotted lines. When the accreting NS comes into the region $\nu>\nu_{\rm c}$, i.e., from point $I$ to $O$ ($\nu_{\rm c}$ along the direction that arrows 3 and 4 point toward), its {\em r}-mode is unstable. {\em Left} and {\em right panels} are for the CEEW and HENW scenarios, respectively. Arrows and their numbers point toward evolution direction. Note that within the time interval between point $A$ and $B$ in the {\em left} panel, $\nu$ and $T$ values do not vary greatly. So these two points nearly overlap. The similar thing also occurs between point $D$ and $O$ in the {\em right} panel.}
	\label{fig:window}
\end{figure*}

Second, the term $A_{ \pm}=1 \pm 3 \alpha^2 \tilde{J} / 2 \tilde{I}$, where $\tilde{J}=\frac{1}{M R^4}\int_0^R \rho r^6 dr=1.635\times10^{-2}$ and $\tilde{I}=\frac{8 \pi}{3 M R^2} \int_0^R \rho r^4 dr=0.261$ for an NS with mass $1.4M_{\odot}$, radius $R=12.5$ km, and polytropic equation of state (EoS) of $N=1$ \cite{owen98,and01}. 

Third, the various timescales involved in Equations (\ref{eq:alpha}) and (\ref{eq:Omega}) are described as follows.
Firstly, $t_{\rm acc}\equiv I\Omega/N_{\rm acc}$ is the accretion timescale, in which the torque is expressed as \cite{piro11}
\be \label{eq:N_acc}
N_{\rm acc}=\begin{cases}\left(1-\frac{\Omega}{\Omega_{\rm K}}\right)(GMR)^{1/2}\dot{M},
	&r_{\rm m}\leqslant R \\
	n(\omega)\left(GM r_{\rm m}\right)^{1/2}\dot{M}, &r_{\rm m}>R,
	\end{cases}
\ee
where $\Omega_{\rm K}=(GM/R^{3})^{1/2}$ is the Keplerian velocity,
$n(\omega)=1-\omega$ is the fiducial dimensionless torque which depends on the
fastness parameter $\omega=\Omega/(GM/r_{\rm m}^3)^{1/2}=(r_{\rm m}/r_{\rm c})^{3/2}$ \cite[e.g.,][]{els77,gho79}.
Secondly, $t_{\rm dip}\equiv I\Omega/N_{\rm dip}$ is the magnetic dipole radiation timescale that reflects the magnetic braking rate associated to the surface magnetic field (here $I=\tilde{I} M R^2$ is the moment of inertia), 
in which the torque for an orthogonal rotator is described by \cite[e.g.,][]{piro11}
\be \label{eq:N_dip}
N_{\rm dip}=-\frac{\mu^2 \Omega^3}{6c^3},
\ee 
where $c$ is the speed of light.
Thirdly, $t_{B, \rm gw}$ represents the timescale\footnote{It's worth noting that the timescale $t_{B, \rm gw}$ in Equations (\ref{eq:alpha}) and (\ref{eq:Omega}), akin to $t_{\rm dip}$, is expected to be negative. This is because the associated GW radiation also spins down the NS.} associated with GW radiation stemming from the deformation induced by the generated toroidal field, is given as \cite{cuo12}
\be \label{eq:t_B_gw}
\frac{1}{t_{B, \rm gw}}= -3.1\times10^{-15} \tilde{I}_{0.261} M_{1.4} R_{12.5}^2 \nu_3^4 \epsilon_{B,-8}^2\ {\rm s^{-1}},
\ee 
where $\nu=\Omega/2\pi=10^3{\rm Hz}~\nu_3$ is the spin frequency of the NS, $\tilde{I}_{0.261}=\tilde{I}/0.261$, $M_{1.4}=M/1.4M_{\odot}$, and $R_{12.5}=R/12.5{\rm km}$. The NS ellipticity $\epsilon_B=10^{-8}~\epsilon_{B,-8}$ is written as \cite{cut02}
\be \label{eq:epsilon}
\epsilon_B= -7.8\times10^{-6} R_{12.5}^4 M_{1.4}^{-2} B_{\rm t,15}^2,
\ee
where  $B_{\rm t,15}=B_{\rm t}/10^{15}{\rm G}$.
Next, $t_{\rm sv}$ and $t_{\rm bv}$ are separately the shear viscosity and bulk viscosity timescales, given by \cite{and01}
\be \label{eq:t_sv}
t_{\rm sv}=2.4 \times 10^8 M_{1.4}^{-5/4} R_{12.5}^{23/4} T_9^2~{\rm s},
\ee
and \cite{owen98}
\begin{equation} \label{eq:t_bv}
	\begin{aligned}
		\frac{1}{t_{\rm bv}}= 7.8 \times 10^{-10} R_{12.5}^3 M_{1.4}^{-1} T_9^6 \nu_3^2~{\rm s^{-1}},
	\end{aligned}
\end{equation}
where $T_9=T/10^9{\rm K}$.
Moreover, $t_{B,\rm t}$ stands for the magnetic damping timescale associated with toroidal field generation, read as \cite{cuo12}
\begin{equation} \label{eq:t_B_t}
	\frac{1}{t_{B,\rm t}} = 1.9\times10^{-3} R_{12.5} M_{1.4}^{-1} B_{12} B_{\rm t,15} \nu_3^{-1}~{\rm s^{-1}}.
\end{equation}
Finally, $t_{\rm gw}$ signifies the GW radiation timescale related to the primary $l = m = 2$ {\em r}-mode, expressed by \cite{lind98,owen98}
\begin{equation} \label{eq:t_gw}
	\begin{aligned}
		\frac{1}{t_{\mathrm{gw}}}= & -\frac{32 \pi G \Omega^6}{225c^7} \left(\frac{4}{3}\right)^6 \int_0^R \rho(r) r^6~dr  \\
		= & -5.26\times10^{-2} M_{1.4} R_{12.5}^4 \nu_3^6\  {\rm s^{-1}}.
	\end{aligned}
\end{equation}

Fourth, the terms $\dot{\epsilon_{\rm s}}$, $\dot{\epsilon_{\rm n}}$, and $\dot{\epsilon_{\nu}}$ in Equation (\ref{eq:T}) determine the global thermal balance of the NS, which separately are the rates of shear viscosity heating associated with electron-electron scattering \cite{and01}, accretion heating due to the accretion compression and pycnonuclear reactions in the NS crust \cite{bro98}, and modified URCA cooling in the NS core \cite{sha83}, given by \cite{and01,bro98}
\begin{equation} \label{eq:eps_s}
	\begin{aligned}
		\dot{\boldsymbol{\epsilon}}_{\rm s} & =\frac{2 \tilde{J}M R^2 \alpha^2 \Omega^2}{t_{\rm sv}}  \\
		& =3.6 \times 10^{37} \tilde{J} M_{1.4}^{9/4} R_{10}^{-15/4} \alpha^2 \Omega^2 T_9^{-2}~{\rm erg~s^{-1}},
	\end{aligned}
\end{equation}
\begin{equation} \label{eq:eps_n}
	\dot{\epsilon}_{\rm n}=\frac{\dot{M}}{m_{\rm n}} \times 1.5~{\rm MeV}=4 \times 10^{51} \dot{M}_{1.4}~{\rm erg~s^{-1}},
\end{equation}
and \cite{sha83}
\begin{equation} \label{eq:eps_nu}
	\dot{\epsilon_{\nu}}=7.5 \times 10^{39} M_{1.4}^{2 / 3} T_9^8~{\rm erg~s^{-1}}.
\end{equation}
Above which $m_{\rm n}$ is the mass of a nucleon \cite{bro98} and $\dot{M}_{1.4}=\dot{M}/1.4M_{\odot}{\rm s^{-1}}$.
The heat capacity $C_{\rm V}$ is given as \cite{wat02}
\begin{equation} \label{eq:C_V}
	C_{\rm V}=1.6 \times 10^{39} M_{1.4}^{1 / 3} T_9~{\rm erg~K}^{-1} .
\end{equation}
By incorporating Equations (\ref{eq:M}) and (\ref{eq:eps_s})-(\ref{eq:C_V}), Equation (\ref{eq:T}) transforms into
\begin{equation} \label{eq:TT}
	\frac{dT}{dt}=\frac{\dot{\epsilon_{\rm s}} + \dot{\epsilon_{\rm n}} - \dot{\epsilon_{\nu}}}{C_{\rm V}} - \frac{T\dot{M}}{6M}.
\end{equation}

In addition to the evolution of the {\em r}-mode instability of the accreting NS in an NS-WD merger, the {\em r}-mode instability window, delineated by the region $\nu>\nu_{\rm c}$ in the contour map between $\nu$ and $T$, also should be known.
The critical frequency $\nu_{\rm c}$ above which the {\em r}-mode becomes unstable \cite{owen98}, can be obtained by solving for the zeros of \cite[see, e.g.,][]{and01,wang17}
\begin{equation} \label{eq:window}
	\frac{1}{-2 E_{\rm c}}\frac{dE_{\rm c}}{dt}=\frac{1}{t_{\rm gw}}+\frac{1}{t_{\rm sv}}+\frac{1}{t_{\rm bv}}+\frac{1}{t_{B, \rm t}}=0,
\end{equation}
where $E_{\rm c}=\frac{1}{2} \tilde{J} M R^2 \alpha^2  \Omega^2$ is the canonical energy of the {\em r}-mode. 

\section{Numerical Calculations} \label{sec:numerical}
\begin{figure}
	\includegraphics[scale=0.48]{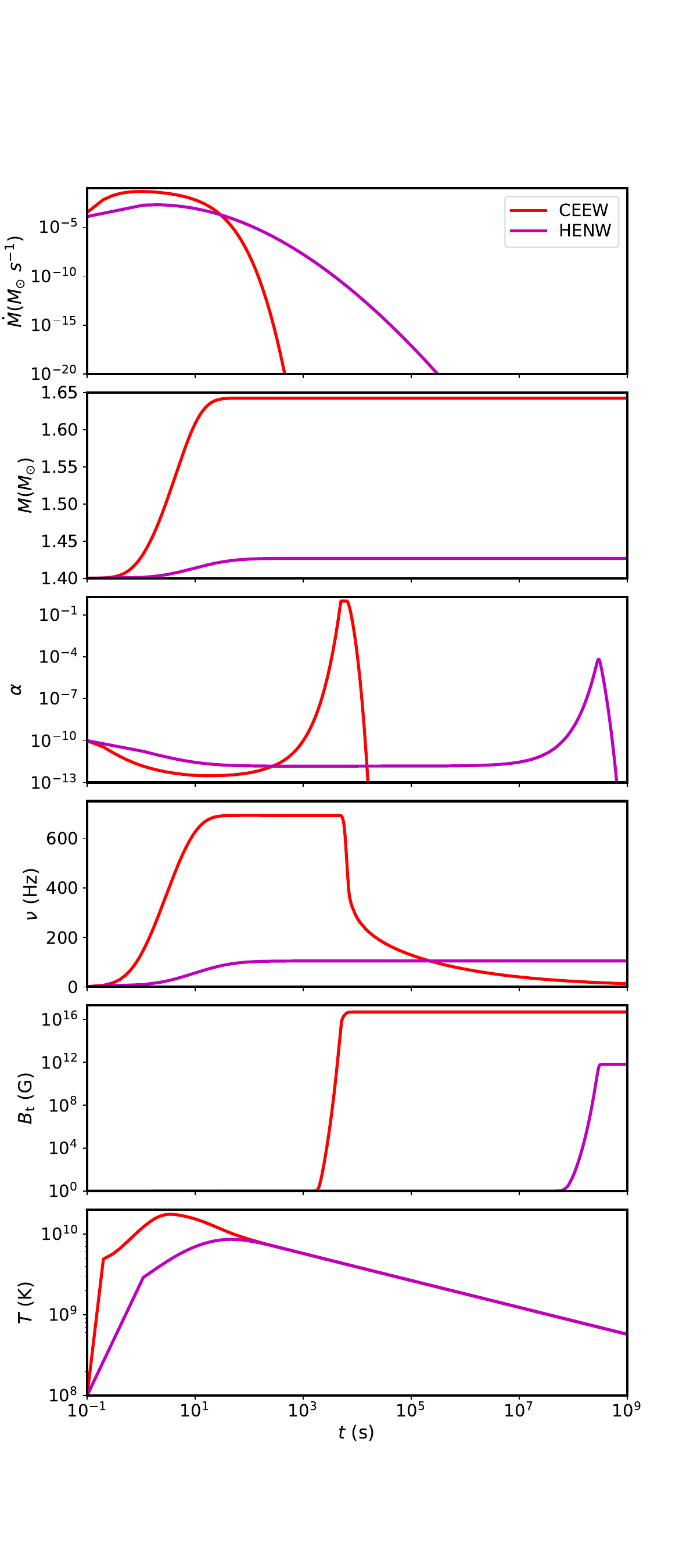}
	\caption{The evolution of the accretion rate $\dot{M}$, mass $M$, {\em r}-mode amplitude $\alpha$, spin frequency $\nu$, volume-averaged toroidal magnetic field $B_{\rm t}$, and spatially averaged temperature $T$ of the NS in an NS-WD merger, for both the CEEW and HENW accretion scenarios [see Equation (\ref{eq:M_dot})].  }
\label{fig:evo}
\end{figure}

We conduct numerical calculations based on Equations (\ref{eq:alpha}), (\ref{eq:Omega}), (\ref{eq:B_t}), and (\ref{eq:TT}) to track the evolution of $\alpha$, $\Omega$, $B_{\rm t}$, and $T$ of the accreting NS in an NS$-$WD merger. The initial parameter values are set as $\alpha_{\rm i}=10^{-10}$, $\Omega_{\rm i}=2\pi$ rad, $B_{\rm t, i}=1$ G, 
$T_{\rm i}=10^8$ K, $M_{\rm i}=1.4M_{\odot}$, $R=12.5$ km, and $B=10^{10}$ G.
The evolution can be divided into distinct phases:
(i) Initially, it is in the $r_{\rm m}\leqslant R$ phase before point $A$ in Figure \ref{fig:window}, the accretion torque $N_{\rm acc}$ is govern by the first line of Equation (\ref{eq:N_acc}). 
At this phase, the evolution comes into the {\em r}-mode instability window where $\nu>\nu_{\rm c}$ from point $I$.
(ii) Subsequently, as the system transitions into the $R<r_{\rm m}\leqslant r_{\rm c}$ phase from point $A$ to $B$\footnote{Note that within the time interval between point $A$ and $B$ in the left panel of Figure \ref{fig:window} for the CEEW scenario, $\nu$ and $T$ values do not vary greatly. So these two points nearly overlap. The similar thing also occurs between point $D$ and $O$ in the right panel of Figure \ref{fig:window} for the HENW scenario.}, the accretion torque enters the column accretion phase, represented by the second line of Equation (\ref{eq:N_acc}).
(iii) Upon entering the $r_{\rm m}>r_{\rm c}$ phase after point $B$, known as the propeller phase, the NS mass stabilizes at $M(t_{\rm c})$ and no longer changes in the subsequent evolution, see Equation (\ref{eq:M}).
(iv) As time progresses, the accretion rate expressed by Equation (\ref{eq:M_dot}) becomes extremely low\footnote{When $t\gtrsim10^4$ s, the accretion rate described by Equation (\ref{eq:M_dot}) becomes extremely low. However, the accuracy of this exceedingly low accretion rate is uncertain due to the absence of numerical simulations for an NS-WD merger beyond $10^4$ s at present, see numerical simulations conducted by, e.g., Refs. \cite{fer19,kal23}.} at point $C$, the terms relevant to accretion in Equations (\ref{eq:alpha}), (\ref{eq:Omega}), (\ref{eq:B_t}), and (\ref{eq:TT}) (i.e., $\dot{M}$, $t_{\rm acc}$, and $\dot{\epsilon_{\rm n}}$) can be thrown out. We regard this time as the end of the accretion.
(v) After the accretion, the {\em r}-mode instability exponentially grows, and the {\em r}-mode amplitude $\alpha$ quickly increases up to a peak at point $D$. If the amplitude $\alpha$ reaches up to ${\cal{O}}(1)$ at which the non-linear effects can no longer be ignored, just like the CEEW scenario in Figure \ref{fig:evo}, one can follow the approach of Ref. \cite{owen98} and then continue to evolve parameters by Equations (\ref{eq:alpha}), (\ref{eq:Omega}), (\ref{eq:B_t}), and (\ref{eq:TT}), but fix $\alpha=1$. 
(vi) Until $d\alpha/dt$ becomes negative because of the gradual dominance of viscosity and toroidal field dampings ($t_{\rm sv}$, $t_{\rm bv}$, and $t_{B,\rm t}$) over the $r$-mode GW term ($t_{\rm gw}$), we free the value of $\alpha$ and continue to evolve parameters by Equations (\ref{eq:alpha}), (\ref{eq:Omega}), (\ref{eq:B_t}), and (\ref{eq:TT}). 
(vii) Ultimately, because the GW radiation from $B_{\rm t}$-induced deformation continues to carry away the angular momentum of the NS and the critical frequency $\nu_{\rm c}$ rises again, the NS falls into $\nu<\nu_{\rm c}$ and the evolution exits the region of $r$-mode instability, depicted by point $O$ in Figure \ref{fig:window}. Note that in Figure \ref{fig:window}, the solid lines represent the evolution of the accreting NS in the system, while the dashed dotted lines represent the evolution of the critical frequency $\nu_{\rm c}$ [obtained by solving Equation (\ref{eq:window})] above which the {\em r}-mode becomes unstable, arrows and their numbers point toward evolution direction.

\begin{figure*}
	\includegraphics[scale=0.43]{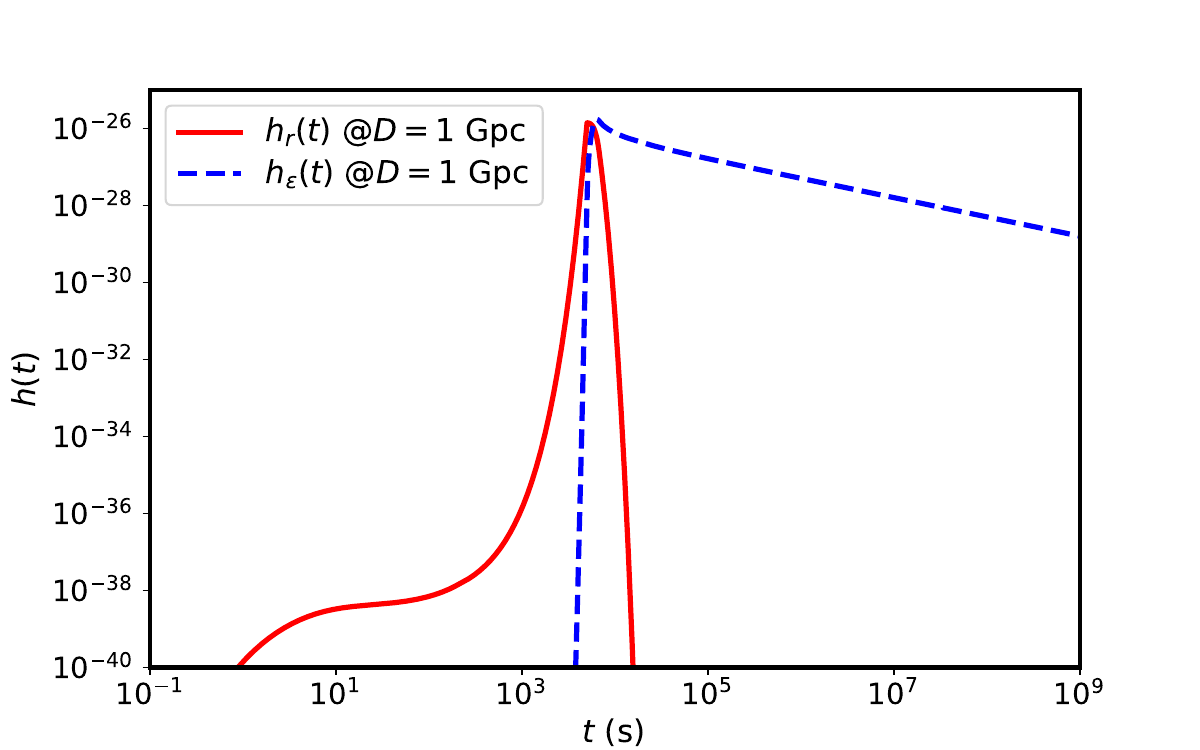}
	\includegraphics[scale=0.43]{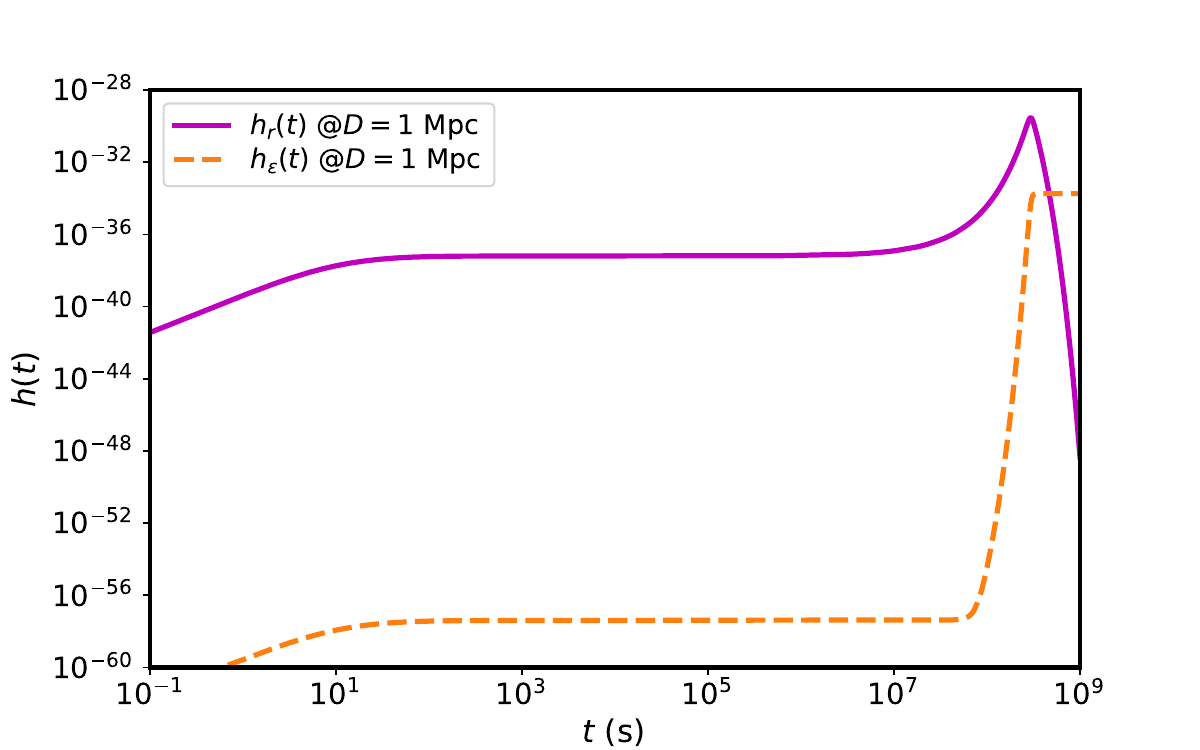}
	\includegraphics[scale=0.43]{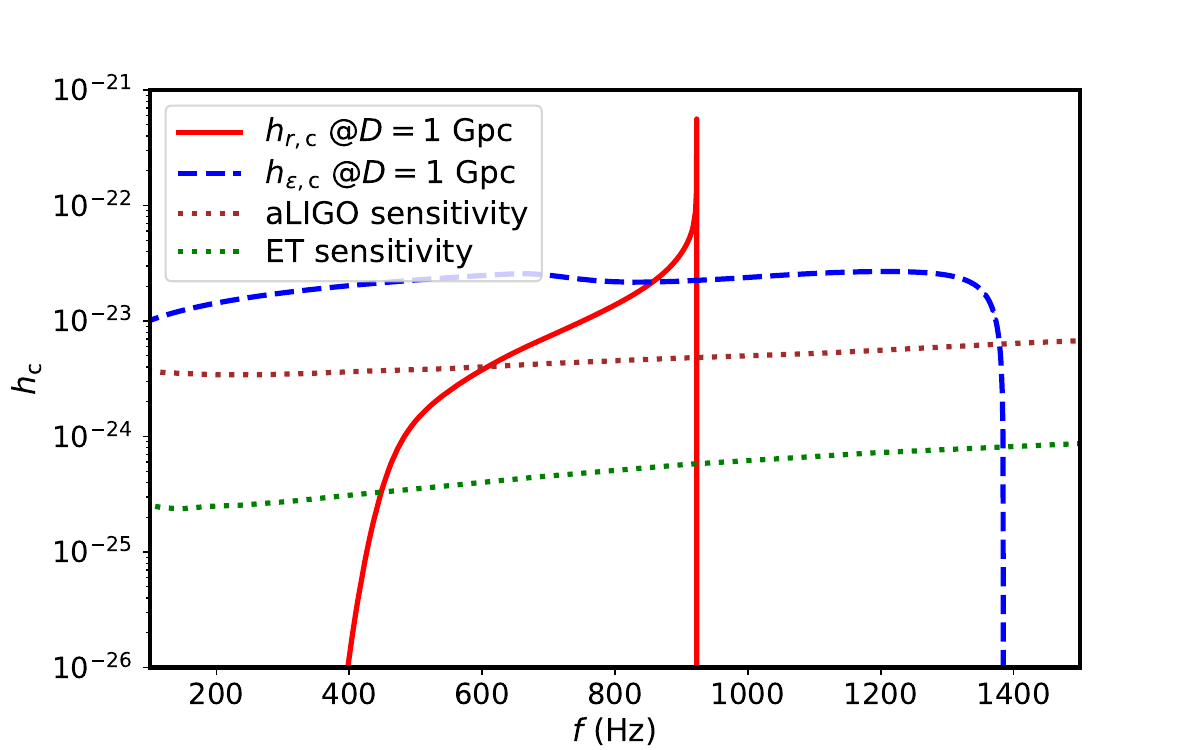}
	\includegraphics[scale=0.43]{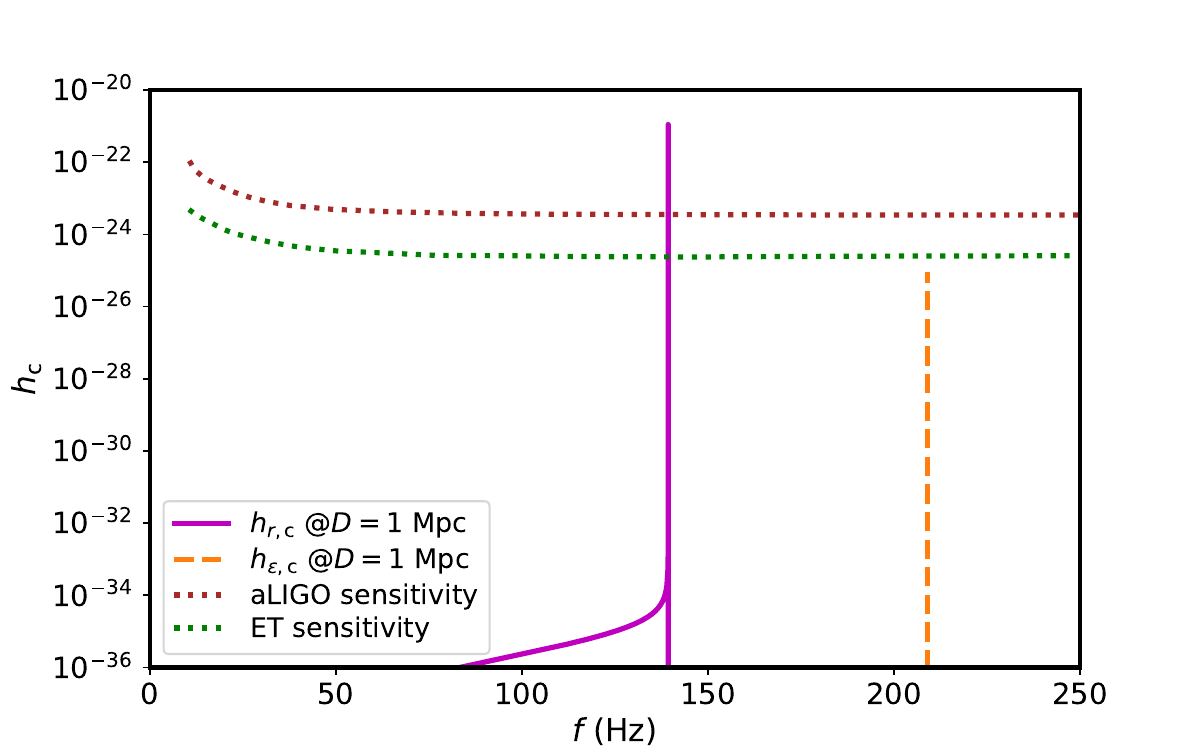}
	\caption{The strain amplitudes related to the {\em r}-mode GW ($h_r$) and $B_{\rm t}$-induced GW ($h_{\epsilon}$) varying with time are shown in the {\em upper panels}. Their respective characteristic amplitudes ($h_{r,\rm c}$ and $h_{\epsilon,\rm c}$) are depicted against the GW frequencies ($f_r$ and $f_{\epsilon}$) in the {\em lower panels}. Results are presented for both the CEEW ({\em left panels}) and HENW ({\em right panels}) accretion scenarios. 
	The calculations for the CEEW scenario are done at a distance of $D=1$ Gpc, while for the HENW scenario, the distance is $D=1$ Mpc. 
	Sensitivities of aLIGO and ET are represented by brown and green dotted lines, respectively.}
	\label{fig:h}
\end{figure*}

The evolution of key parameters including the accretion rate $\dot{M}$, mass $M$, {\em r}-mode amplitude $\alpha$, spin frequency $\nu$, volume-averaged toroidal magnetic field $B_{\rm t}$, and spatially averaged temperature $T$ of the NS are depicted in Figure \ref{fig:evo}, contrasting the outcomes for both the CEEW and HENW accretion scenarios.  
On one hand, for either the CEEW or HENW scenario, its temperature evolution closely tracks the accretion rate, primarily due to the dominance of accretion heating outlined in Equation (\ref{eq:eps_n}). The amplification of $B_{\rm t}$ mainly results from the growth in {\em r}-mode amplitude. For the spin frequency evolution of the NS, its initial rapid increase is owing to the rapid accretion and its final state is determined by the rotational energy loss via GW radiation from the $B_{\rm t}$-induced deformation, as evidenced in the upper panels of Figure \ref{fig:h}. Note that the results illustrated in Figure \ref{fig:h} are based on Section \ref{sec:gw} below.
On the other hand, in the CEEW scenario, a higher peak accretion rate $\sim0.046M_{\odot}{\rm s^{-1}}$ leads to a higher final mass $\sim1.64M_{\odot}$, larger peak {\em r}-mode amplitude ${\cal{O}}(1)$, higher peak spin frequency $\sim690$ Hz, stronger toroidal magnetic field $\sim5\times10^{16}$ G, and elevated peak temperature $\sim1.8\times10^{10}$ K of the final NS, compared to the HENW scenario. 
Specifically, the rapid spin-down behavior of the NS in the CEEW scenario is attributed to the persistent strong GW radiation originating from the magnetar-like $B_{\rm t}$, as illustrated in the left upper panel of Figure \ref{fig:h}. Conversely, the comparatively stable spin in the HENW scenario is owed to the subdued GW radiation resulting from the relatively weak $B_{\rm t}$, as indicated in the right upper panel of Figure \ref{fig:h}.

\section{GW Radiation}\label{sec:gw}
There are two types of GW radiation in the aforementioned NS-WD merger event. One type directly arises from the $r$-mode instability of the NS, primarily the $l = m = 2$ {\em r}-mode. Its strain amplitude averaged over polarizations and orientations can be calculated by \cite{owen98}
\begin{equation} \label{eq:h}
	h_r (t)=\sqrt{\frac{3}{80 \pi}} \frac{(4\Omega/3)^2 S_{22}}{D},
\end{equation}
where
\begin{equation}
	S_{22}=\sqrt{2} \frac{32 \pi}{15} \frac{G M}{c^5}  \tilde{J}R^3 \alpha \Omega,
\end{equation}
is the $l = m = 2$ current multipole moment and $D$ is the distance to the source. 
The characteristic amplitude of the GW signal corresponding to $h_r (t)$, compared to the noise amplitude of GW detectors, is defined as \cite{owen98}
\begin{equation} \label{eq:h_c}
	h_{r, \rm c} \equiv h_r(t) \sqrt{f_r^2\left|\frac{d t}{d f_r}\right|},
\end{equation}
where $f_r=2\Omega/(3\pi)$ is the GW frequency (i.e., the mode frequency) measured in Hz.

The second type is relevant to the NS deformation due to the generation of toroidal magnetic field, indirectly associated with the $r$-mode instability. 
Its strain amplitude can be estimated by \cite[e.g.,][]{cor09}
\begin{equation} \label{eq:h_eps}
	h_{\epsilon}(t)=\frac{4 G I \epsilon_B}{D c^4} \Omega^2.
\end{equation}
While its characteristic amplitude $h_{\epsilon,\rm c}$ can be computed in a similar way as Equation (\ref{eq:h_c}) but with a different GW frequency $f_{\epsilon}=\Omega/\pi$.

From the evolution results outlined in Section \ref{sec:evolution}, one can obtain the temporal and spectral properties of these two types of GW radiation. As exhibited in Figure \ref{fig:h}, 
the CEEW scenario, characterized by a higher peak accretion rate, yields stronger GW signals compared to the HENW scenario. Specifically, for CEEW, both {\em r}-mode and $B_{\rm t}$-induced GWs, spanning durations of $\gtrsim10^3$ s and $\gtrsim10^4$ s, respectively, and frequency ranges of $\sim500-900$ Hz and $\sim100-1400$ Hz, can be detected even at a distance of $\gtrsim10$ Gpc by aLIGO or $\gtrsim100$ Gpc by ET. In contrast, for HENW, its weak $B_{\rm t}$-induced GW cannot be detected at a Mpc distance by aLIGO or ET. While its {\em r}-mode GW, lasting up to $\sim10^8$ s and falling in a narrow frequency $\sim140$ Hz, 
can be detected at a distance $\gtrsim10$ Mpc by aLIGO or $\gtrsim100$ Mpc by ET. 
Obviously, these GWs for both the CEEW and HENW scenarios, are much stronger than those associated with {\em r}-mode instability and magnetic deformation induced by {\em r}-mode instability in NS-WD binaries categorized as IMXBs or LMXBs  \cite{levin99,and99,cuo12,mah13}.

Additionally, the volumetric event rate of NS-WD mergers is estimated to be $\sim90-5800~{\rm Gpc^{-3}~yr^{-1}}$ \cite{too18,kal23,kang24}. 
Given uncertainties regarding the number of mergers involving WDs with mass $\gtrsim$ 1 $M_{\odot}$, under the CEEW scenario, 
the most optimistic projected event numbers for {\em r}-mode and $B_{\rm t}$-induced GWs could reach $\sim(9-580)\times10^4~{\rm yr}^{-1}$ 
for aLIGO or $\sim(9-580)\times10^7~{\rm yr}^{-1}$ for ET. 
As a comparison, under the HENW scenario, the event number for {\em r}-mode GW may peak at $\sim(9-580)\times10^{-5}~{\rm yr}^{-1}$ for aLIGO 
or $\sim(9-580)\times10^{-2}~{\rm yr}^{-1}$ for ET.

In a word, these strong GW emissions and potential high event rate make such NS-WD mergers interesting sources for aLIGO and ET.

\section{EM Counterparts}\label{sec:counterparts}
Some intriguing EM emissions have been linked to an NS-WD merger. Notably, certain peculiar long gamma-ray bursts (LGRBs) have been proposed to create from such NS-WD mergers \cite{king07,cha07}. Recent investigations have reinforced this notion, suggesting that peculiar LGRBs, such as GRBs 211211A \cite{ras22,yang22,mei22,tro22,gom23,zhang22,xiao22,meng24} and 230307A \cite{levan24,sun23,gil23,yang24,dic23}, may be the outcome of NS-WD mergers \cite{yang22,zhong23,zhong24,wang24}, especially when the WD's mass exceeds $\sim1M_{\odot}$ and its debris disk after tidal disruption is in the CEEW scenario with a high accretion rate \cite{zhong23}. If this is the case, the GW emissions explored in this work can be applied for identifying whether such peculiar LGRBs are indeed generated by an NS-WD merger.

Furthermore, fast blue optical transients (FBOTs) including kilonova-like emissions associated with peculiar LGRBs have also been proposed to originate from NS-WD mergers, based on both theoretical modelings for observations \cite{kas10,dro13,mcb19,gil20,chen20,pren20,zhong23} and 
hydrodynamical–thermonuclear numerical simulations \cite{fer19,zen19,zen20,bob22,kal23,mor24}.

Moreover, there's speculation that certain fast radio bursts (FRBs), like FRB 20200120E occurring within a globular cluster \cite{bha21,kir22}, might generate from highly magnetized NSs formed post NS-WD mergers\footnote{It is important to notice that the formation of a remnant with a magnetar-like internal field, creating from an NS-WD merger via the toroidal magnetic field amplification mechanism known as the $\omega$ dynamo induced by differential rotation which is one of three mechanisms discussed in Ref. \cite{zhong20}, requires the NS-WD system to be in the CEEW scenario, rather than the HENW scenario previously used in Ref. \cite{zhong20}. This conclusion is evident, as shown by the comparison for the peak values of $B_{\rm t}$ between these two scenarios in Figure \ref{fig:evo}. } \cite{zhong20,kre21}. 

As a result, the types of GWs discussed in Section \ref{sec:gw} could potentially accompany these EM events, offering valuable insights into the merger process and the progenitor nature.

\section{Summary}\label{sec:summary}
We've analyzed the evolution of key parameters pertinent to the {\em r}-mode instability of the NS in an NS-WD merger with WD's mass $\gtrsim 1M_{\odot}$, including the {\em r}-mode amplitude $\alpha$, spin angular frequency $\Omega$, volume-averaged toroidal magnetic field $B_{\rm t}$, and spatially averaged temperature $T$.
According to the evolution results, we've further studied the related {\em r}-mode and $B_{\rm t}$-induced GWs. 
Our investigation encompasses both the CEEW and HENW prescriptions for the WD debris disk, as detailed in Ref. \cite{kal23}.
Here are the important findings:
\begin{itemize}
\item In the CEEW scenario, characterized by a higher peak accretion rate ($\sim0.046M_{\odot}{\rm s^{-1}}$), one can observe substantial differences compared to the HENW scenario. The CEEW leads to a higher final mass ($\sim1.64M_{\odot}$), larger peak {\em r}-mode amplitude (${\cal{O}}(1)$), higher peak spin frequency ($\sim690$ Hz), stronger toroidal magnetic field ($\sim5\times10^{16}$ G), and elevated peak temperature ($\sim1.8\times10^{10}$ K) of the final NS.
\item The temporal and spectral characteristics of both the {\em r}-mode GW and $B_{\rm t}$-induced GW exhibit distinct features between the CEEW and HENW scenarios. 
Notably, the CEEW scenario yields much stronger GW signals compared to the HENW scenario. Specifically, for CEEW, both {\em r}-mode and $B_{\rm t}$-induced GWs can be detected even at a distance of $\gtrsim10$ Gpc by aLIGO or $\gtrsim100$ Gpc by ET. On the contrary, for HENW, its weak $B_{\rm t}$-induced GW cannot be detected at a Mpc distance by aLIGO or ET. While its {\em r}-mode GW
can be detected at a distance $\gtrsim10$ Mpc by aLIGO or $\gtrsim100$ Mpc by ET. 
Furthermore, these GWs for both the CEEW and HENW scenarios, 
are much stronger than those associated with {\em r}-mode instability in NS-WD binaries categorized as IMXBs or LMXBs \cite{levin99,and99,cuo12,mah13}.
Moreover, the most optimistic projected event numbers for {\em r}-mode and $B_{\rm t}$-induced GWs could be considerably large for the CEEW scenario, $\sim(9-580)\times10^4~{\rm yr}^{-1}$ for aLIGO or $\sim(9-580)\times10^7~{\rm yr}^{-1}$ for ET, as compared to those in the HENW scenario. 
Accordingly, such strong GW emissions and potential high event rate make such NS-WD mergers interesting sources for aLIGO and ET.
\item These {\em r}-mode and $B_{\rm t}$-induced GWs could be associated with certain peculiar LGRBs, FBOTs containing kilonova-like emissions associated with peculiar LGRBs, and/or FRBs, offering valuable insights into the merger dynamics and the nature of the progenitor system.
\end{itemize}

Some caveats should be mentioned in this work. First, for simplicity we have used a traditional and commonly-used treatment of the NS with a polytropic EoS of $N=1$. A more realistic EoS and modifications for traditional {\em r}-mode instability window such as the stellar matter nonbarotropicity and relativity, as well as the efficiency of particle diffusion in damping the relativistic r-modes \cite{kraa24} should be considered. This will be studied in our future work. Second, the NS-WD merger systems in our study only have a short dynamical accretion timescale with the order of magnitude of $10^3$ s for CEEW scenario or $10^4$ s for HENW scenario. Moreover, once the merger finishes, i.e., the accretion ends and disk wind escapes, only the single NS leaves behind. Therefore, for NS-WD merger systems, there is only a one-off relatively short evolution as illustrated in Figure \ref{fig:window}. This is not similar to the long-term cyclic evolution in LMXBs as shown in the figure 1 of Ref. \cite{levin99}.

\acknowledgements We thank Zi-Gao Dai and Long Li for helpful discussions. 
This work is supported by the starting Foundation of Guangxi University of Science and Technology (grant No. 24Z17).
Y.Z.M. is supported by the starting Foundation of Guangxi University of Science and Technology(grant No. 24Z01)

%\bibliography{ms}

\end{document}